\documentclass{pasj00}
\draft

\begin{document}
\SetRunningHead{Yang \& Meng}{Frequency separation variations}
%\Received{ }%{yyyy/mm/dd}
%\Accepted{ }%{yyyy/mm/dd}

\title{Frequency separation variations of the solar low-degree p-modes}

%%% begin:list of authors
% Do NOT capitalize all letters in "textsc".
\author{Wuming Yang}
 \and
\author{Xiangcun Meng}
\affil{School of Physics and Chemistry, Henan Polytechnic
University, Jiaozuo 454000, Henan, China.}
\email{wuming.yang@hotmail.com}

%\author{C-Firstname {\sc C-Familyname}}
%\affil{C-Address of Institute}\email{ccccc@xxx.xxx.xx.xx}
%%% end:list of authors

%%% Please use the following style in case that sorting by
%%% affilation is impossible.
%
% \author{%
%   D-Firstname \textsc{D-Familyname}\altaffilmark{1}
%   E-Firstname \textsc{E-Familyname}\altaffilmark{1,2}
%   and
%   F-Firstname \textsc{F-Familyname}\altaffilmark{2}}
% \altaffiltext{1}{Address of Institute}
% \email{ddddd@xxx.xxx.xx.xx}
% \email{eeeee@xxx.xxx.xx.xx}
% \altaffiltext{2}{Address of Institute}

%% `\KeyWords{}' always has to be placed before `\maketitle'.
 %Do NOT move this preamble from here!

\maketitle

\begin{abstract}
Variations of frequency separations of low-degree p-modes are
studied over the solar activity cycle. The separations studied are
obtained from the frequencies of low-degree p-modes of the Global
Oscillation Network Group (GONG). 10.7 cm radio flux is used as an
index of solar activity. Small separations of the p-mode frequencies
are considered to be mainly dependent on the conditions in stellar
interiors. Thus they could be applied to diagnose the changes in the
stellar interior. Our calculation results show that the magnitudes
of variations of the mean large separations are less than 1 $\sigma$
over the solar activity cycle. Small separations show different
behaviors in the ascending and descending phases of activity. In the
ascending phase, variations of the small separations are less than 1
$\sigma$. However, the small separations have systematic shifts
during 2004 - 2007. The shifts are roughly 1 $\sigma$ or more. The
variations of the ratios of the small to large separations with time
are similar to the changes of the small separations. The effects of
the changes in the large separations on the ratios are negligible.
The variations of the separations may be a consequence of the
influence from the surface activity or systematic errors in
measurements or some processes taking place in the solar interior.

\end{abstract}

\section{Introduction}
It is well-known that the solar p-mode frequencies vary with the
solar cycle \citep{wood85, elsw90, libb90, elsw94, jime98, howe99,
chap07}. It has been shown that the variations of the frequencies
are significantly related to magnetic activities near the solar
surface \citep{wood91, jime04, tou05}. However, a number of attempts
have been made to detect the relation between the frequency shifts
and the variation of the solar interior. It has been proposed that
the frequency shifts may relate to changes taking place in the base
of the convection zone \citep{jime98, sere05}, or in the deep core
\citep{jime98}.

Low-degree p-modes can be used to study the solar core because they
can penetrate the deep interior \citep{dzie97}. However, even these
modes are primarily sensitive to the solar envelope structure. Thus
frequency separations are proposed for a diagnostic tool. The
frequency separations of low-degree p-modes in solar-like stars
often proposed for diagnostic purposes are large separations
$\Delta_{l}(n)$ and small separations $d_{ll+2}(n)$. The large
separations approximate to the characteristic frequency $\nu_{0}$ of
a star, defined by
\begin{equation}
  \Delta _{l}(n)\equiv \nu_{n, l} - \nu_{n-1, l} \simeq \nu_{0},
  \label{eqla}
\end{equation}
where
\begin{equation}
  \nu_{0} = (2\int^{R}_{0}dr/c)^{-1},
\end{equation}
in which $c$ is sound speed at radius $r$ and $R$ is some fiducial
radius of the star. The small separations are sensitive to the
structure of stellar core \citep{gou90a}, defined by
\begin{equation}
\begin{array}{lll}
   d_{l l+2}(n)& \equiv & \nu_{n, l} - \nu_{n-1, l+2}\\
   & =& \frac{\varphi_{l+2}-\varphi_{l}}{\pi}\nu_{0},
\end{array}
   \label{eqsm}
\end{equation}
where $\varphi_{l}$ is the internal phase shift. For the low-degree
p-modes, the $\varphi_{l}$ is mainly dependent on conditions in the
stellar core and is almost independent of those outside the core
\citep{rox03}. These frequency separations have been extensively
investigated by many authors \citep{chris84, ulr86, ulr88, gou87,
gou90b, gou03, gou90a, rox03, aud94, rox05, flo05}. Additionally
\citet{rox93} defines differences $d_{01}(n)$ as
\begin{equation}
\begin{array}{lll}
   d_{01}(n)& \equiv & \frac{-\nu_{n, 1} + 2\nu_{n, 0}- \nu_{n-1,
   1}}{2} \\
  & = & \frac{\varphi_{1}-\varphi_{0}}{\pi}\nu_{0},
\end{array}
\end{equation}
and \citet{yang07} define differences $\sigma_{l-1l+1}(n)$ of
low-degree p-modes as
\begin{equation}
\begin{array}{lll}
  \sigma_{l-1 l+1}(n)& \equiv & -\nu_{n, l-1}+2\nu_{n, l} - \nu_{n,
  l+1} \\
  & = & \frac{\varphi_{l+1}+ \varphi_{l-1}-2\varphi_{l}}{\pi}\nu_{0}.
\end{array}
\end{equation}
The differences $\sigma_{02}(n)$ are similar to the scaled small
separations $d_{l l+2}(n)/(2l+3)$ in some cases and are mainly
sensitive to conditions in stellar core. The difference
$\sigma_{02}$ averaged over $n$ is, however, more sensitive to
changes in the central hydrogen abundance of solar-like stars than
the scaled small separations \citep{yang07}.

Moreover, the ratios $d_{l l+2}/\Delta_{l}$ are considered to be
essentially independent of the structure of the outer layers of a
star and determined by interior structures \citep{ulr86, rox03,
rox05, flo05}. The ratios $d_{01}/\Delta_{0}$ and
$\sigma_{02}/\Delta_{0}$ also mainly rely on the interior structures
\citep{rox03, yang07}. Thus changes in stellar interior could be
indicated by the variations in these separation ratios.

However frequency variations with solar activity are relative to the
angular ($l$) degree, the azimuthal ($m$) degree, and the order $n$.
The $m$-dependence is attributed to the solar near-surface activity
\citep{jime04, tou05} and could be removed by using the frequency
centroids which incorporate all the $m$ components of a mode
\citep{chap05}. For the low-degree p-modes, centroid shifts can be
regarded as being $l$-independent \citep{chap05}. Thus separations
of frequency centroids of low-degree p-modes should be insensitive
to effects of surface activity. Using the p-modes of the Birmingham
Solar-Oscillations Network (BiSON) collecting data using a
Sun-as-a-star technique, \citet{chap05} studied the impact of the
solar activity cycle on the frequency separation ratios and found
that the ratios change with the shifting level of global solar
activity. \citet{chap05} pointed out that some $m$ components are
suppressed in the Sun-as-a-star observations and hence the ratios
are expected to be affected by the surface activity. In this paper,
we investigate the variations of frequency separations of low-degree
p-modes as solar activity cycle proceeds using data collected by the
resolved-Sun observations. In Section 2 we present our data and
results. Then discussion and conclusion are represented in Section
3.

\section{Data and results}
In this work, we used the solar p-mode frequencies of the GONG
\citep{harv96} to study the separations. The GONG was planned for
measurements of medium and high spherical degrees, but low-degree
p-modes can be measured too. \citet{gavr99} showed that the low-l
p-modes of the GONG are in good agreement with that of the Michelson
Doppler Imager (MDI) and the Global Oscillations at Low Frequency
(GOLF). The raw data of GONG are measured in a time series of 36
days labeled a GONG month. A 108-day time series is constructed by
concatenating 3 consecutive GONG-month time series. GONG Month Mode
Frequencies (GMFs) were then estimated from the 108-day spectra
\citep{ande90, hill98, komm99, howe99}. In the observation of GONG,
the solar disk is imaged onto many pixels. The obtained images can
then be decomposed into their constituent spherical harmonics. For
each degree $l$ and overtone $n$, this imaging strategy provides
access to all $2l+1$ components with each being tagged by an
azimuthal degree $m$ \citep{chap04}.

The frequencies, $\nu_{n,l,m}$, extracted from resolved-Sun
observations are usually represented by
\begin{equation}
  \nu_{n,l,m}=\nu^{c}_{n,l}+\sum^{2l}_{j=1}a_{j}(n,l)l\mathcal{P}^{j}_{l}(m),
  \label{num}
\end{equation}
where $\mathcal{P}^{j}_{l}(m)$ are polynomials related to
Clebsch-Gordan coefficients \citep{ritz91, chap04} and
$\nu^{c}_{n,l}$ is the so-called central frequency of the multiplet
(i.e. frequency centroid). Using the polynomials
$\mathcal{P}^{j}_{l}(m)$ given by \citet{chap04} and equation
(\ref{num}), one can get the central frequency of low-degree p-modes
of the resolved-Sun observations,
\begin{equation}
  \nu^{c}_{n,l}=\sum^{l}_{m=-l}\frac{\nu_{n,l,m}}{2l+1}.
  \label{centf}
\end{equation}
The central frequencies are sensitive to the spherically symmetric
component of the internal structure of a star and are usually used
as inputs to inversions of the internal structure \citep{chap04,
chap05}.

\begin{figure}
  \begin{center}
  \includegraphics[angle=-90,scale=0.4]{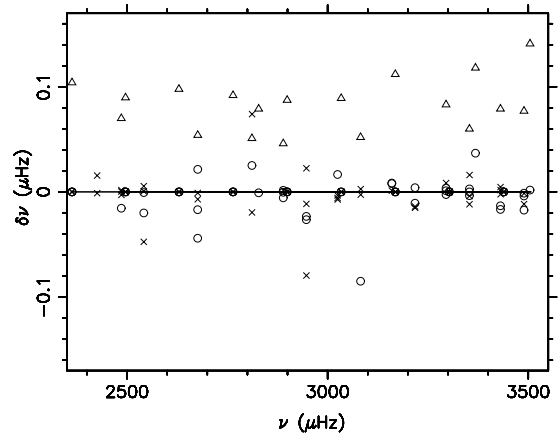}
  \end{center}
\caption{Differences between the averages obtained using equation
(\ref{centf}) and the centroids estimated by the GONG team. Circles
represent the differences of frequencies of the GONG months 8, 9,
and 10 (low activity), while crosses correspond to the differences
of frequencies of the GONG months 58, 59, and 60 (high activity).
Triangles show the errors of GONG centroids of month 8.}
\label{fig1}
\end{figure}

In order to study the variations of frequency separations with the
solar activity cycle, we used the GMFs observed between 1996 and
2007. Centroids $\nu_{n,l}$ and individual frequencies $\nu_{n,l,m}$
of the GMFs were published by the GONG
team\footnote{ftp://gong2.nso.edu/MFS/}. Some centroids $\nu_{n,l}$
published are inconsecutive in the order $n$, especially for the
modes with $l$ = 1, but the frequencies $\nu_{n,l,m}$ are
consecutive. In order to obtain separations, we used the averages
$\nu^{c}_{n,l}$ of all $2l$ + 1 components of each mode instead of
the centroids $\nu_{n,l}$. In figure \ref{fig1}, we compared the
averages obtained using equation (\ref{centf}) with the available
centroids estimated by the GONG team. The differences between the
frequencies obtained using equation (\ref{centf}) and the available
GONG centroids are almost much less than the errors of the GONG
centroids in the range of 2400 $\mu$Hz $< \nu <$ 3500 $\mu$Hz
\textbf{except a few modes, whose differences are close to the
errors of the GONG centroids}. In this range, the averages obtained
using equation (\ref{centf}) are good consistent with the centroids
estimated by the GONG team. All the following frequency separations
are obtained from the averages.

\begin{figure}
  \begin{center}
  \includegraphics[angle=-90,scale=0.35]{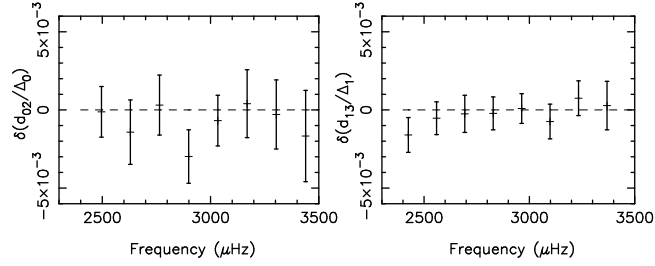}
  \end{center}
\caption{ Fractional changes between the separations ratios in two
GMFs, one at a high level (2001) and one at a low level of activity
(1996) (the differences are in the sense "high data - low data").}
\label{fig2}
\end{figure}

Figure \ref{fig2} shows fractional changes of the separation ratios
in two sets of GMFs, one at a low level and one at a high level of
activity. The error bars indicate 1 $\sigma$ errors, which were
propagated from uncertainties on the individual frequencies of the
GMFs. The departure of the separation ratios from the null level is
within 1 $\sigma$ but shows a systematic shift, which is consistent
with the prediction of the 108 day spectra of \cite{chap05} (in
their figure 1). However, just as the observed shifts of ratios of
the BiSON frequencies \citep{chap05}, `the shifts are clearly not
significant for individual estimates of the ratios'. In the interest
of temporal variations of separations, for each set of separations
with the same degree $l$, we averaged the separations over $n$ in
the range of 2400 $\mu$Hz $< \nu_{n,l} <$ 3500 $\mu$Hz, which was
implied by symbols $< >$. The frequency range is often used to
obtain a mean frequency shift \citep{elsw94, jime04}.

\begin{figure}
\begin{center}
\includegraphics[angle=-90,scale=0.4]{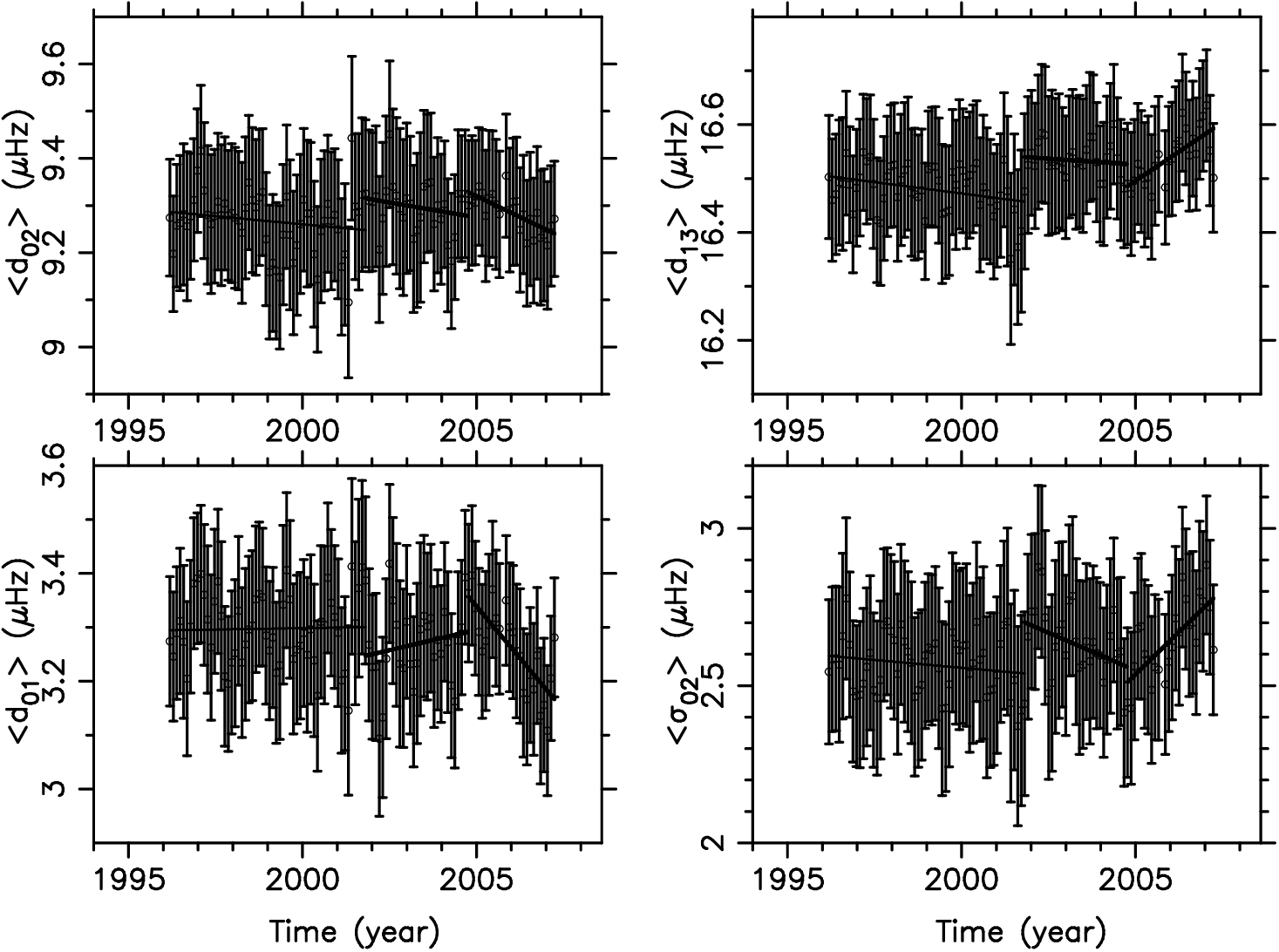}
\includegraphics[angle=-90,scale=0.4]{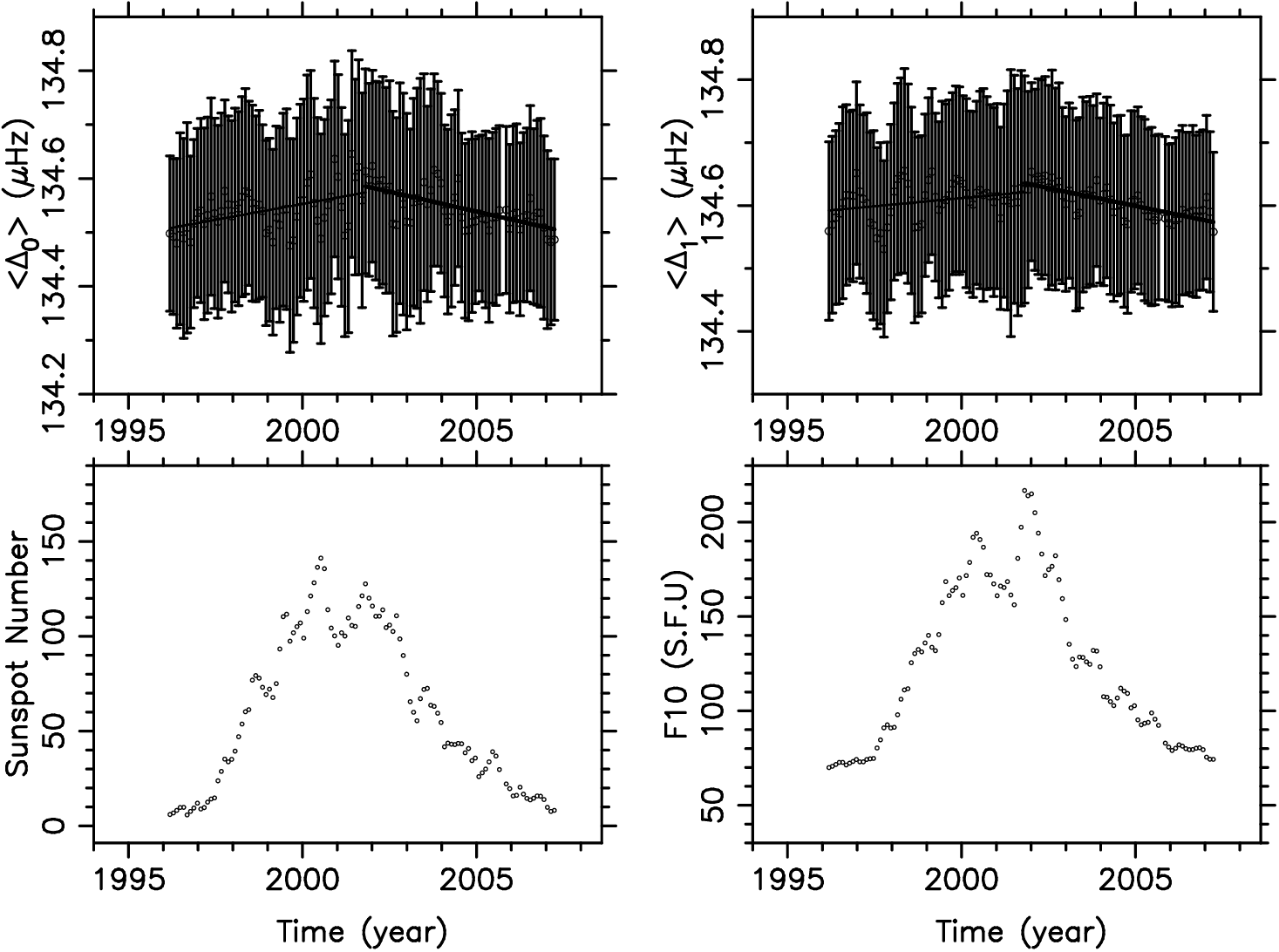}
\end{center}
\caption{Temporal variation of separations. The error bars represent
1 $\sigma$ errors. The thin lines show the linear fits between the
plotted quantities in the ascending phase of the solar activity,
while the thick lines indicate the linear fits in the descending
phase of the solar activity. Sunspot number and 10.7 cm radio flux
are plotted too. 1 SFU = 10$^{-22}$ W $\cdot$ m$^{2}$ $\cdot$
Hz$^{-1}$.} \label{fig3}
\end{figure}

%------------------------------------------------------
   \begin{table}
   \begin{center}
   \caption{Results of linear fits between separations and time.
   Left panel for the ascending phase of solar activity, while right
   panel for the during $\sim$ 2004 to 2007 except for the large separations.
   \label{tab1}}
   \begin{tabular}{cccc|ccc}
   \hline \hline
          & Ascending phase&  &  & \multicolumn{3}{c}{From $\sim$ 2004 to 2007} \\
   \hline
    Separation  & Slope (nHz/year)$^{a}$ & $\chi^{2}$ & R & Slope (nHz/year)$^{a}$ & $\chi^{2}$ & R \\
   \hline
     $<d_{02}>$      & - 6.9 $\pm$ 5.1 & 0.228 & 0.177 & - 35.4 $\pm$ 7.6 & 0.019 & 0.697 \\
     $<d_{01}>$      & 1.1 $\pm$ 5.3  & 0.244 & 0.027  & - 76.9 $\pm$ 13.4 & 0.058 & 0.768 \\
     $<\sigma_{02}>$ & - 9.9 $\pm$ 7.1 & 0.446 & 0.184 & 108 $\pm$ 23 & 0.179 & 0.692 \\
     $<d_{13}>$      & - 8.6 $\pm$ 3.3 & 0.092 & 0.327 & 43.0 $\pm$ 9.0 & 0.029 & 0.691 \\
     $<\Delta_{0}>$  & 11.8 $\pm$ 2.8  & 0.072 & 0.485 & - 14.7 $\pm$ 2.3$^{b}$ & 0.041 & 0.654 \\
     $<\Delta_{1}>$  & 5.2 $\pm$ 2.2   & 0.041 & 0.303 & - 11.4 $\pm$ 1.3$^{b}$ & 0.012 & 0.771 \\
   \hline
     $<d_{02}/3\Delta_{0}>$     &- (1.9 $\pm$ 1.2)$\times 10^{-5}$& 1.3$\times 10^{-6}$ & 0.205
     & - (8.8 $\pm$ 1.9)$\times 10^{-5}$ & 1.2$\times 10^{-7}$ & 0.698 \\
     $<d_{01}/\Delta_{0}>$      &(5.6 $\pm$ 3.9)$\times 10^{-5}$& 1.3$\times 10^{-5}$ & 0.019
     & - (5.7 $\pm$ 1.0)$\times 10^{-4}$ & 3.2$\times 10^{-6}$ & 0.766 \\
     $<\sigma_{02}/\Delta_{0}>$ &- (7.5 $\pm$ 5.3)$\times 10^{-6}$& 2.5$\times 10^{-5}$  &0.188
     & (8.0 $\pm$ 1.7)$\times 10^{-4}$ &9.8 $\times 10^{-6}$ & 0.694 \\
     $<d_{13}/5\Delta_{1}>$     &- (1.4 $\pm$ 0.5)$\times 10^{-5}$& 2.0$\times 10^{-7}$ &0.348
     & (6.5 $\pm$ 1.4)$\times 10^{-5}$ & 6.2 $\times 10^{-8}$ & 0.699 \\
   \hline
   \end{tabular}
\begin{list}{}{}
\item a. Not for ratios.
\item b. The results were obtained from the data between $\sim$ 2001 and 2007.
\end{list}
\end{center}
\end{table}
%---------------------------------------------------------------

Figure \ref{fig3} shows the temporal variation of mean separations
$<d_{ll+2}>$, $<\sigma_{02}>$, $<d_{01}>$, $<\Delta_{0}>$ and
$<\Delta_{1}>$. Sunspot numbers and 10.7 cm radio flux
\footnote{www.ngdc.noaa.gov/stp/SOLAR/getdata.html} (F10), which are
always used as an index of the solar activity, are also plotted as a
function of time. The activity indices have been averaged over the
same time interval covered by the GONG date. \textbf{The error bars
show 1 $\sigma$ errors, which come from the formal errors on the
frequencies. When averaging these separations from each spectrum,
the correlations of neighbouring separations were considered}. The
lines in this figure represent the results of linear least-squares
fits. The fits in the descending phase of solar activity were
divided into two periods in accordance with the variations of
separations: one from $\sim$ 2001 to 2004 and one from $\sim$ 2004
to 2007. Variations in the latter period are more obvious and
systematic. The mean large separations $<\Delta_{0}>$ and
$<\Delta_{1}>$ increase/decrease in the ascending/descending phase
of solar activity and reach a maximum in 2001. However, the
magnitude of the shifts of the large separations is about 0.07
$\mu$Hz, which is less than 1 $\sigma$. Variations of small
separations with time are more complicated than those of the large
separations. In the ascending phase of solar activity, the
separations $<d_{ll+2}>$, $<\sigma_{02}>$ and $<d_{01}>$ almost have
no systematic shifts; the magnitudes of the shifts of these
separations, obtained from their regression equations, are also less
than 1 $\sigma$. In the descending phase of solar activity,
variations of these separations seem to be disorderly. Between 2001
and 2002, the separations $<\sigma_{02}>$ and $<d_{13}>$ increase
suddenly. They decrease firstly and then increase obviously between
$\sim$ 2001 and 2007. However, the magnitudes of their shifts are
roughly 1 $\sigma$ or more. The separation $<d_{01}>$ has an obvious
and systematic decrease between $\sim$ 2004 and 2007. The magnitude
of the variation is about 1.5 $\sigma$. The separation $<d_{02}>$
decreases slightly between $\sim$ 2001 and 2007, but the magnitude
is less than 1 $\sigma$. The results of linear fits: slope, residual
sum of squares $\chi^{2}$, and correlation coefficient $R$ of
regression equations, are summarized in Table \ref{tab1}.

\begin{figure}
  \begin{center}
  \includegraphics[angle=-90,scale=0.35]{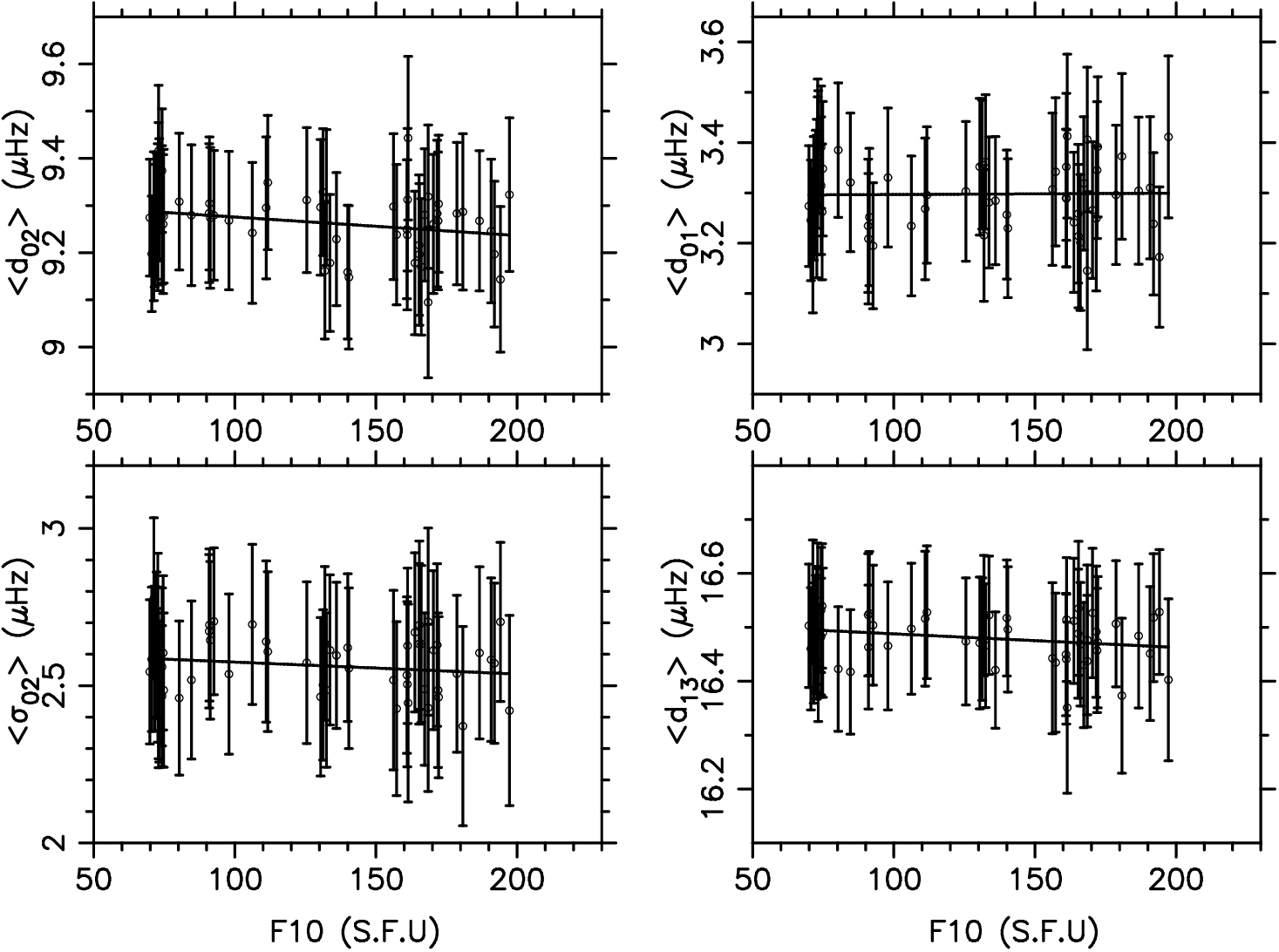}
  \includegraphics[angle=-90,scale=0.35]{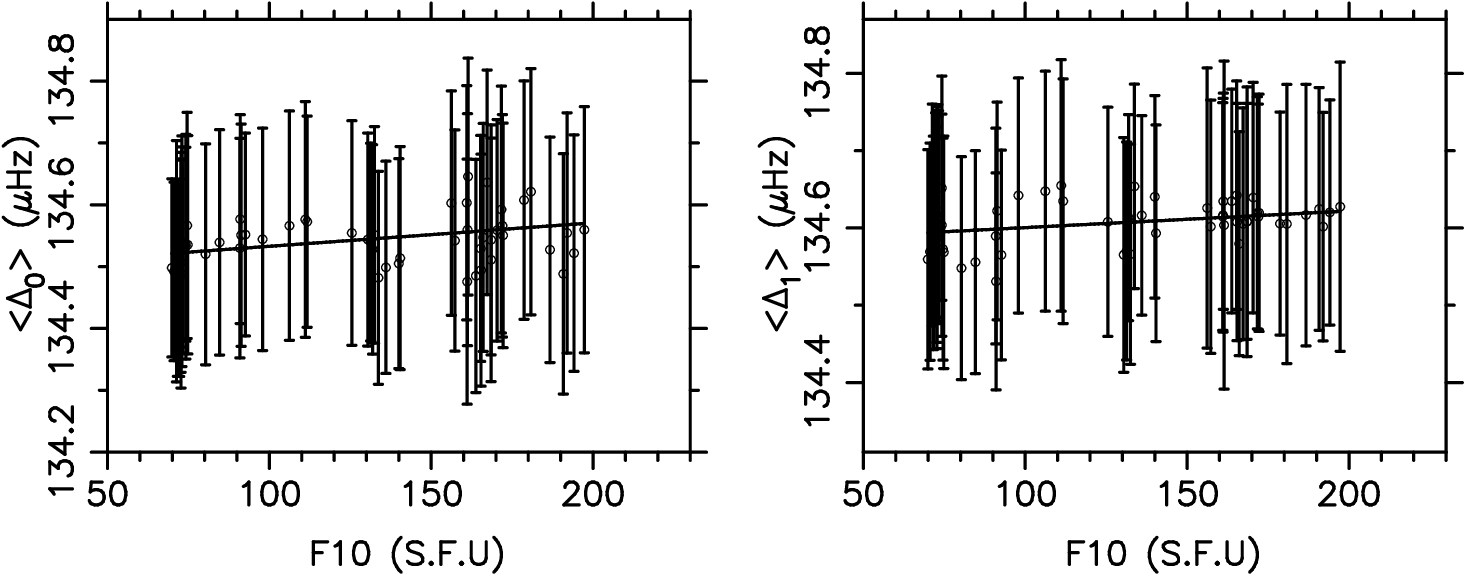}
  \end{center}
  \caption{Mean separations are plotted as a function of 10.7 cm radio flux
  in the ascending phase of solar activity.
  The error bars indicate 1 $\sigma$ errors. The lines represent the best
  linear fits between the plotted quantities.} \label{fig4}
\end{figure}

\begin{figure}
  \begin{center}
  \includegraphics[angle=-90,scale=0.35]{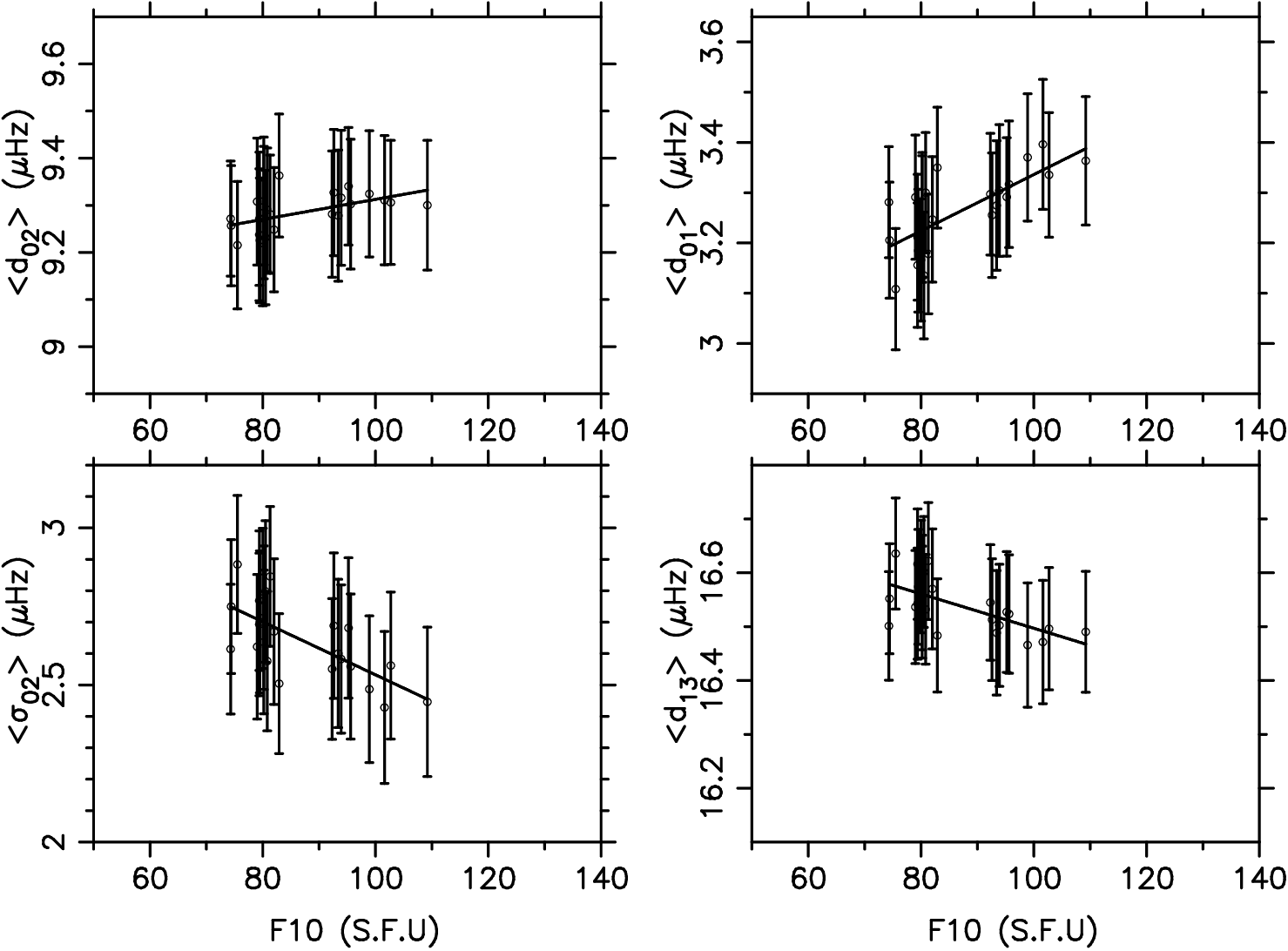}
  \includegraphics[angle=-90,scale=0.35]{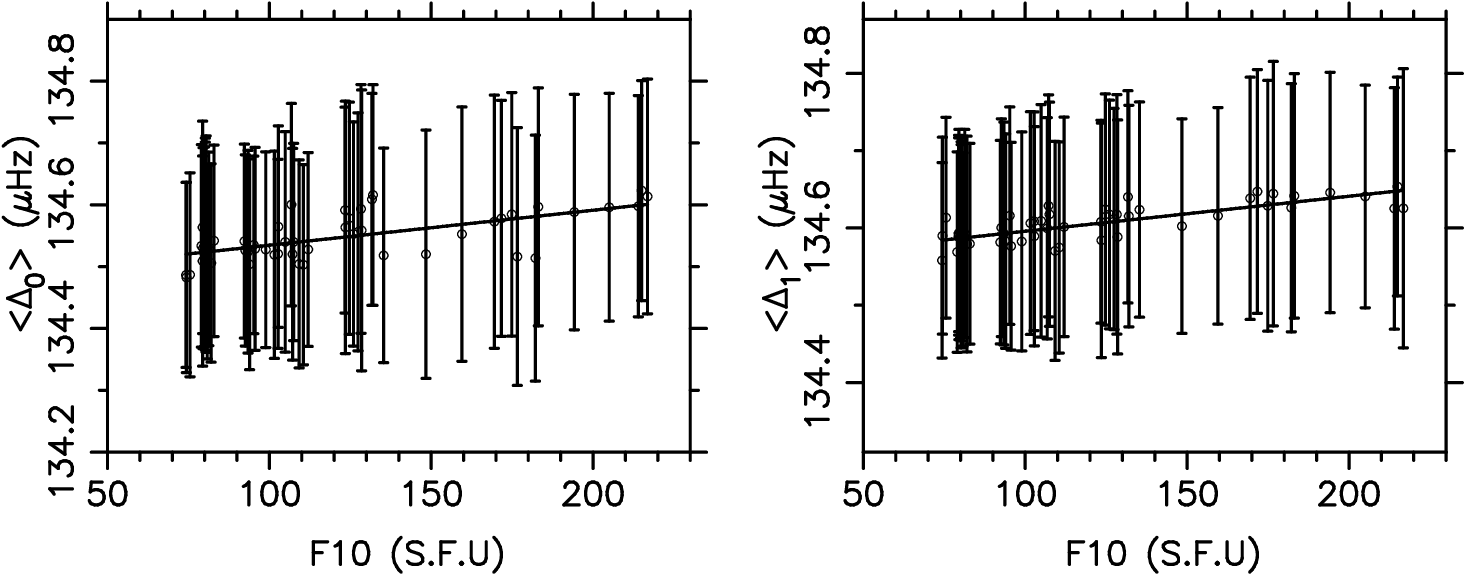}
  \end{center}
  \caption{Mean separations observed between $\sim$ 2004 and 2007 are plotted
  as a function of 10.7 cm radio flux. But the large separations are
  observed between $\sim$ 2001 and 2007.
  The error bars indicate 1 $\sigma$ errors. The lines represent the best
  linear fits between the plotted quantities.} \label{fig5}
\end{figure}
%------------------------------------------------------
   \begin{table}
   \begin{center}
   \caption{Results of linear fitting between separations and 10.7 cm
   radio flux.
   \label{tab2}}
   \begin{tabular}{cccc|ccc}
   \hline \hline
          & Ascending phase&  &  & \multicolumn{3}{c}{From $\sim$ 2004 to 2007} \\
   \hline
    Separation  & Slope (nHz/SFU) & $\chi^{2}$ & R & Slope (nHz/SFU) & $\chi^{2}$ & R \\
   \hline
     $<d_{02}>$      & - 0.34 $\pm$ 0.19 & 0.223 & 0.235 & 2.13 $\pm$ 0.69 & 0.026 & 0.542 \\
     $<d_{01}>$      & 0.024 $\pm$ 0.197 & 0.244 & 0.016 & 5.6 $\pm$ 1.1 & 0.068 & 0.722 \\
     $<\sigma_{02}>$ & - 0.33 $\pm$ 0.27 & 0.449 & 0.162 & - 8.4 $\pm$ 1.8 & 0.176 & 0.699 \\
     $<d_{13}>$      & - 0.25 $\pm$ 0.13 & 0.096 & 0.254 & - 3.2 $\pm$ 0.8 & 0.032 & 0.657 \\
     $<\Delta_{0}>$  & 0.33 $\pm$ 0.11   & 0.081 & 0.361 & 0.57 $\pm$ 0.09$^{a}$ & 0.041 & 0.648 \\
     $<\Delta_{1}>$  & 0.22 $\pm$ 0.08   & 0.040 & 0.329 & 0.45 $\pm$ 0.05$^{a}$ & 0.012 & 0.791 \\
   \hline
   \end{tabular}
\begin{list}{}{}
\item a. The results were obtained from the data between $\sim$ 2001 and 2007.
\end{list}
\end{center}
\end{table}
%---------------------------------------------------------------

In order to further understand the variations of separations with
the solar activity, Figure \ref{fig4} shows separations plotted
against F10 in the ascending phase of solar activity. Figure
\ref{fig4} shows that the separations almost not vary with activity.
The results of linear least-squares fits between the separations and
F10 are summarized in the left panel of Table \ref{tab2}. The
magnitudes of variations of these separations are only about 0.04
$\mu$Hz, which is less than 1 $\sigma$. Moreover, Figure \ref{fig5}
shows that the changes of the large separations with the F10 in the
descending phase of the solar activity. The shifts of the large
separations are also less than 1 $\sigma$. Figure \ref{fig5} also
shows the variations of small separations with the F10 during $\sim$
2004 - 2007. In this period of time, the changes of these
separations with the F10 are obvious and systematic. However, the
magnitudes of variations of these separations are around 1 $\sigma$
or more. The results of linear fits between the separations and F10
are shown in the right panel of Table \ref{tab2}.

\begin{figure}
  \begin{center}
  \includegraphics[angle=-90,scale=0.35]{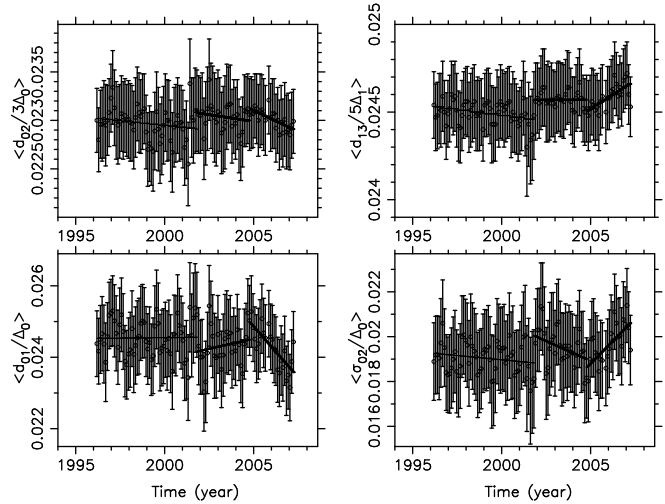}
  \end{center}
\caption{Temporal variations of separation ratios. The error bars
indicate 1 $\sigma$ errors. The lines show the linear fits between
the plotted quantities.} \label{fig6}
\end{figure}

In Figure \ref{fig6}, we show the temporal variations of separation
ratios $<d_{02}/3\Delta_{0}>$, $<d_{13}/5\Delta_{1}>$,
$<d_{01}/\Delta_{0}>$ and $<\sigma_{02}/\Delta_{0}>$. The results of
linear fits between the ratios and time are presented in Table
\ref{tab1}. The changes of the ratios with time are similar to those
of the corresponding small separations. In the ascending phase of
solar activity, shifts of the ratios are less than 1 $\sigma$. In
the descending phase of activity, from $\sim$ 2004 to 2007, the
changes of the ratios with time are systematic. However, the
magnitudes of the changes are still roughly 1 $\sigma$ or more.
Since the fractional changes of the large separations are much less
than those of small separations during the activity, the
contribution to the changes in the ratios from the variations of the
large separations is much less than that from the small separations.
The effects of the changes in the large separations on the changes
in the ratios are almost negligible. The changes in the ratios with
time are mainly determined by the changes in the small separations.

The magnitudes of temporal variations for some individual
separations are very large but should not be significant. In the
ascending phase of the solar activity, the magnitudes of variations
of separations, obtained from regression equations, are about 0.04
$\mu$Hz, which is less than 1 $\sigma$. During $\sim$ 2004 - 2007,
the changes in separations are systematic. However, the magnitudes
of the changes are roughly 1 $\sigma$ or more. Using BiSON data,
\citet{chap05} show that variations of the separation ratios between
low and high activity is at a marginal level of significance.
Moreover, using simple averages, from the gradient of mean frequency
shift of the MDI given by \citet{jime04} in their Table 1, one can
get a shift gradient in the centroids would be 1.61 and 2.02 nHz/SFU
for $l$ = 0 and 2 respectively. For a 120 SFU increase in solar
activity, the change in separation $<d_{02}>$ would then be about
0.05 $\mu$Hz, which is consistent with our result.

\section{Discussion and conclusion }

The characteristic frequency $\nu_{0}$, i.e. the large separation,
is an integral property of the entire star. Because the solar sound
speed is a decreasing function of the solar radius, the contribution
to $\nu_{0}$ from the outer layers of the Sun is larger than that
from the solar inner regions \citep{gou90a}. Thus the large
separation should be more sensitive to the changes in the outer
layers than those in the inner regions and could be affected by the
near surface activity. Moreover small separations are more sensitive
to the conditions in stellar interior than those in outer layers.
Thus the large separations should be more sensitive to the activity
than the small separations. However, in fact, the magnitudes of
variations of the large separations are about 0.07 $\mu$Hz during
the activity, which is less than 1 $\sigma$. The magnitude of
variations of $<d_{02}>$ is also less than 1 $\sigma$ over the
activity cycle. However, other small separations show a systematic
shift during $\sim$ 2004 - 2007, and the magnitudes of their
variations are roughly 1 $\sigma$ or more.

Small separations $<d_{ll+2}>$, $<\sigma_{02}>$ and $<d_{01}>$ show
a sudden change between 2001 and 2002. The fractional changes shown
in Figure \ref{fig2} of ratios in two GMFs (one in 1996 and one in
2001) are consistent with the prediction of \citet{chap05}. However,
if we use the data of 2002 instead of the data of 2001, the results
would be different. The small separations show different behaviors
in the ascending and descending phases of the cycle. In the
ascending phase, variations of the small separations are less than 1
$\sigma$. In the descending phase, the small separations show a
systematic shift during 2004 - 2007. And the magnitudes of the
changes in $<d_{13}>$, $<\sigma_{02}>$ and $<d_{01}>$ are roughly 1
$\sigma$ or more. It is not clear whether the variations are caused
by the systematic errors in measurements.

The large separations increase slightly with increasing of the solar
activity. But the magnitudes of temporal variations of the mean
large separations of the low-degree p-modes are only about 0.07
$\mu$Hz over the activity cycle, which is less than 1 $\sigma$.
Small separations $<d_{ll+2}>$, $<\sigma_{02}>$ and $<d_{01}>$ show
different behaviors in the ascending and descending phases of the
cycle. In the ascending phase, variations of the small separations
are less than 1 $\sigma$. However, the small separations show a
systematic shift during 2004 - 2007. The magnitudes of the changes
in $<d_{01}>$, $<\sigma_{02}>$ and $<d_{13}>$ in this period of time
are about 1 $\sigma$ or more. The variations of the ratios of the
small to large separations with time are similar to the changes of
the small separations. Changes in the large separations have a
negligible impact on the variations in the ratios. The variations of
the separations may be a consequence of the influence from the
surface activity or systematic errors in measurements or some
processes taking place in the solar interior. For further work, the
data over cycle 22/23 would be required.

\section*{Acknowledgments}
This work utilizes data obtained by the Global Oscillation Network
Group (GONG) program, managed by the National Solar Observatory,
which is operated by AURA, Inc. under a cooperative agreement with
the National Science Foundation. The data were acquired by
instruments operated by the Big Bear Solar Observatory, High
Altitude Observatory, Learmonth Solar Observatory, Udaipur Solar
Observatory, Instituto de Astrofisica de Canarias, and Cerro Tololo
Interamerican Observatory.

\end{document}